# Outlier-Robust unscented Kalman filter based on generalized correntropy induced

**JINHUI HU**

Southwest Jiaotong University, Chengdu, China

**HAIQUAN ZHAO**, Senior Member, IEEE

Southwest Jiaotong University, Chengdu, China

**YI PENG**

Southwest Jiaotong University, Chengdu, China

***Abstract*—Conventional Kalman filtering (KF) approaches exhibit significant limitations in addressing nonlinear state estimation problems contaminated by non-Gaussian noise disturbances. To overcome these challenges, this work proposes a robust iterative square root unscented Kalman Filter based on the generalized correntropy induced (SR-GCI-IUKF). While sharing the maximum correntropy criterion's (MCC) ability to characterize higher-order noise statistics, the proposed GCI framework exhibits intrinsic kernel bandwidth insensitivity—a critical advantage enabling robust adaptation to diverse complex noise environments through its generalized kernel structure. For nonlinear state estimation challenges, the algorithm constructs a nonlinear error generalization model that dynamically corrects measurement-induced errors during the state update phase, thereby significantly enhancing estimation accuracy in strongly nonlinear regimes. Furthermore, the square-root decomposition implementation ensures numerical robustness by preserving covariance matrix positive definiteness throughout recursive operations. Theoretical stability guarantees are established through rigorous error dynamics analysis, demonstrating bounded estimation variance under non-Gaussian disturbances. Finally, experiments are carried out in nonlinear systems, land vehicle navigation systems as well as power system FASE to compare other robust algorithms, and it is determined that the proposed algorithm has stronger robustness.**

## I. INTRODUCTION

The Kalman filtering (KF) is widely used for state estimation of linear Gaussian systems [1]. To address nonlinear systems, extensions of the KF such as the Extended Kalman Filter (EKF) [2] and the Unscented Kalman Filter (UKF) [3] have been developed. EKF uses first order Taylor expansion to linearize the nonlinear system to achieve the problem of state estimation in nonlinear systems. On the basis, IEKF [5] was developed to further improve the performance of EKF in facing nonlinear systems. The UKF, on the other hand, employs the unscented transform to approximate the probability distribution function, which has higher accuracy in facing nonlinear systems compared to the EKF. And, the iterative UKF [3] is also proposed in an iterative form, which further improves the ability of the UKF algorithm to cope with nonlinear state estimation problems.

The conventional KF mentioned earlier employs the minimum mean square error (MMSE) to address noise issues in state estimation. However, it leads to performance degradation of the KF when confronted with non-Gaussian noise environments. Therefore, KF faces challenges in maintaining optimal performance, resulting in a decline in the accuracy of state estimation [3]. To address this issue, several algorithms combining EKF, UKF, and CKF with different information-theoretic learning methods have been developed. Among them, the maximum correntropy criterion (MCC) and minimum error entropy (MEE) demonstrate powerful performance in handling non-Gaussian noise problems. Combining with the above algorithms, such as MCC-EKF [8] and MEE-UKF [9] were developed for non-Gaussian problem processing in nonlinear systems. And the IMCC-UKF [10] algorithm is developed by combining the MCC criterion with the nonlinear iterative framework, which shows more powerful performance in facing nonlinear systems. And in some cases, due to the low flexibility of ordinary Gaussian kernel, it is difficult to deal with complex noise environments, so the generalized Gaussian kernel is proposed, the algorithms such as GMC-CKF [11], GMEE-UKF [12], and so on, are further developed. Furthermore, combining Gaussian kernels with varying kernel widths enhance algorithm performance, as demonstrated by GMMC-UKF [13] algorithms are developed by it. As well as using MEE and MCC for hybridization, a Minimum Error Entropy with Fiducial Points (MEEF) [14] algorithm was developed, and a MEEF-CKF [15] algorithm was proposed in combination with CKF. However, correntropy-based Kalman filtering is sensitive to the choice of kernel bandwidth; smaller kernel widths can improve the robustness of the algorithm but may lead to filtering dispersion, and larger kernel widths can speed up convergence but affect the accuracy of the filtering [16][17]. Therefore, choosing the appropriate bandwidth is a difficult task for correntropy-based KF [17]. And a correntropy induced (CI) has the ability to capture higher order noise statistics similar to the maximum correlation entropy for higher order noise statistics [18][19], and has a similar treatment of noise as the maximum likelihood estimation, which makes it insensitive to the choice of kernel

This work was partially supported by National Natural Science Foundation of China (grant: 62171388, 61871461, 61571374) and Major Special Project of China Railway Group Limited (Source -2025- Special Project -001).

Jinhui Hu, Haiquan Zhao and Yi Peng are with the Key Laboratory of Magnetic Suspension Technology and Maglev Vehicle, Ministry of Education, School of Electrical Engineering Southwest Jiaotong University Chengdu, China. (jhhu_swjtu@126.com; hqzhao_swjtu@126.com; pengyi1007@163.com) (*Corresponding author: Haiquan Zhao*) width. Specifically, the generalized correntropy induced (GCI) [19] is more robust than the traditional CI, which is largely due to its greater flexibility in parameter selection and the ability to adapt to different noise types. Although CI and GCI both use kernel-based loss functions to suppress outliers, GCI introduces additional tuning parameters, usually including kernel bandwidth and shape parameters, to better describe the complex noise distribution. Unlike CI, which usually depends on a fixed Gaussian kernel, GCI usually uses a more flexible generalized Gaussian kernel, whose shape and tail behavior can be adjusted by parameters. This makes GCI more effective in modeling heavy tail noise, impulse interference and other non-Gaussian characteristics. By adjusting these parameters, GCI achieves a better balance between anomaly suppression and detail preservation in various noise environments.

In addition, another major problem with the correntropy-based UKF algorithm is its numerical stability. It should be noted that the correntropy-based UKF algorithm involves Cholesky decomposition of the error covariance matrix when updating the error covariance matrix. Therefore, it is necessary to maintain the semi-positive characterization of the error covariance matrix, otherwise it will lead to an interruption in the iterative process of the algorithm. A common solution is the square root algorithms, which effectively ensures the positive definiteness of the error covariance matrix during the algorithm update process [20][21]. While square-root MC-KF [22] of square root MC-CKF [23] effectively address this by preserving positive definiteness, they do not solve the deeper limitations inherent to the standard correntropy framework itself. Specifically, the practical adoption of these filters is hindered by three persistent challenges: (1) pronounced sensitivity to kernel bandwidth selection, forcing a fragile trade-off between outlier rejection and filter stability; (2) lack of an inherent mechanism to correct errors introduced by strong nonlinearities during state updates; and (3) limited flexibility of the standard Gaussian kernel in modeling complex, real-world noise distributions such as asymmetric, heavy-tailed, or multimodal. Although square-root implementations address numerical stability, they inherit the core limitations of the underlying correntropy criterion. To overcome these specific shortcomings, this paper proposes a Robust Iterative Square-Root Unscented Kalman Filter based on the Generalized Correntropy Induced metric (SR-GCI-IUKF). The distinct innovations are threefold: First, the GCI framework is employed, whose generalized kernel structure exhibits intrinsic insensitivity to bandwidth selection, eliminating the critical yet cumbersome parameter tuning required by prior methods. Second, a nonlinear error generalization model is constructed to explicitly account for and dynamically correct measurement-induced errors in strongly nonlinear regimes, a feature absent in existing correntropy-based filters. Third, the algorithm integrates square-root decomposition for covariance updates within this robust iterative framework, ensuring numerical stability while simultaneously solving the aforementioned theoretical and practical limitations. The main contribution of the article is as below

1. The SR-GCI-IUKF algorithm was developed based on GCI to overcome the limitations of traditional KF in non-Gaussian noise and strongly non-linear systems. Its generalized kernel structure makes the algorithm insensitive to kernel bandwidth, significantly improving its adaptability to complex noise environments. Furthermore, by constructing a nonlinear error generalization model, it dynamically corrects errors introduced by strong nonlinearities during the state update phase, significantly enhancing estimation accuracy in nonlinear systems.

2. An automated parameter adaptation strategy is designed, and a systematic computational complexity analysis of the algorithm is provided. To address the parameter tuning challenge for the shape and scale parameters in GCI, an adaptive selection strategy based on information potential matching is proposed. Concurrently, a detailed analysis of the computational burden associated with key steps is presented.

3. Through rigorous error dynamics analysis, it is proven that the estimation error of SR-GCI-IUKF remains bounded in the mean square sense under non-Gaussian disturbances. This theoretically ensures the algorithm's reliable convergence and numerical stability in complex noise and highly nonlinear scenarios.

To demonstrate the comprehensive advantages of the proposed algorithm more clearly, this paper systematically compares SR-GCI-IUKF with existing mainstream robust filtering algorithms from four key dimensions: noise resistance, nonlinear processing capability, parameter sensitivity, and computational complexity. As shown in Table I, SR-GCI-IUKF, through its generalized correlation entropy framework and nonlinear error generalization model, comprehensively solves the inherent limitations of traditional methods in bandwidth sensitivity and nonlinear correction while maintaining numerical stability.

The rest of the article is organized as below. The generalized correntropy induced criterion is presented in Section II, a detailed derivation of the algorithm is carried out and error convergence analysis is given in Section III, followed by simulation results in Section IV, and conclusions are drawn in Section V.

## II. Generalized Correntropy Induced

Correntropy is a measure of similarity between two random variables. Assuming two random variables X, Y, the CI is defined as

$$\mathrm{CI}(\mathrm{X},\mathrm{Y}) = \frac{\sigma^2}{2N}\left(1-\sum_{i=1}^{N}\exp\left(-\left(\frac{e_i}{\sigma}\right)^2\right)\right) \tag{1}$$

where $e_i = \mathrm{X}_i - \mathrm{Y}_i$, $\sigma$ is the kernel width of the Correntropy induced. Due to the limitation of the Gaussian kernel, it is difficult to exert sufficient performance in special cases. Therefore, a generalized Gaussian kernel is used instead of the Gaussian kernel. The GCI cost function is constructed and defined as follows [19]

$$\mathrm{GCI}(\mathrm{X},\mathrm{Y}) = \left[\frac{\mu}{N}\left(1-\sum_{i=1}^{N}\exp\left(-\varphi\left|e_i\right|^{\delta}\right)\right)\right]^{1/\delta} \tag{2}$$

TABLE I
COMPARISON OF ROBUST KALMAN FILTERING ALGORITHMS

| Algorithm | Noise Resistance | Nonlinearity Handling | Parameter Sensitivity | Computational Complexity |
|---|---|---|---|---|
| MCC-UKF | Strong against light-tailed non-Gaussian noise | Standard UT transform; no dynamic error correction | High (highly sensitive to kernel bandwidth) | Lower |
| MEE-UKF | Good for various entropy-based noise distributions | Standard UT transform; no dynamic error correction | Moderate (depends on kernel width) | Higher |
| SR-CI-IUKF | Robust to impulsive noise | Iterative update; limited nonlinear correction | Low (insensitive to kernel bandwidth) | Lower |
| SR-GCI-IUKF | Superior in heavy-tailed, mixed, and asymmetric noise | Nonlinear error generalization model; dynamic correction | Very Low (insensitive to tuning of shape & scale) | High |
| SR-GCI-IUKF (optimization) | Best overall robustness | Same as above | Automatically adapted | Highest |

where $\mu = \delta / 2\theta\,\Gamma(1/\delta)$ is the normalization constant, $\Gamma(\bullet)$ is the Gamma function, $\delta > 0$ is the shape parameter, $\theta > 0$ is the scale parameter and $\varphi = 1/\theta$ is the kernel factor. If Y = 0 and with $i \to \infty$, $|e_i| = |\mathrm{X}_i| > 0$, $\forall i : e_i \neq 0$, then a good approximation is obtained by vanilla. When $\delta = 2$, GCI will degenerate to CI. Therefore, the form of the GCI cost function used in this paper as follows

$$\| e \| \sim \mathrm{GCI}^{\delta}(e) = \frac{\mu}{N}\left(1 - \sum_{i=1}^{N} \exp\left(-\varphi |e_i|^{\delta}\right)\right) \tag{3}$$

**Remark:** GCI possesses the same form of entropy as the maximum correntropy, so it can effectively capture higher order moments of the error and thus effectively deal with the effects of non-Gaussian noise. However, unlike the stringency of MCC's kernel bandwidth requirement, GCI is insensitive to kernel bandwidth. The value of kernel bandwidth does not have an excessive impact on the stability of GCI and its performance on noise processing. Moreover, GCI has the property of generalized Gaussian kernel, which has more flexible kernel shapes in dealing with all kinds of noise environments, helping it to maintain good performance in all kinds of environments. Furthermore, compared with CI, which depends on a fixed Gaussian kernel, GCI employs a generalized Gaussian kernel that offers a more flexible shape profile. This structural advantage enables GCI to adapt to a broader range of noise types, including heavy-tailed, impulsive, and mixed non-Gaussian disturbances. Consequently, GCI maintains robust performance across diverse noise environments without requiring intensive parameter tuning. This property renders it particularly advantageous over CI in practical scenarios where noise characteristics are unknown or time-varying.

## III. ROBUST ITERATIVE SQUARE ROOT UKF BASED ON GCI

This subsection outlines the derivation of SR-GCI-IUKF and establishes its stability through convergence analysis of the estimation error.

### A. Robust iterative square root UKF based on GCI

Consider the following nonlinear dynamic system

$$\mathbf{x}_{\tau} = \mathbf{f}\left(\mathbf{x}_{\tau-1}\right) + \mathbf{w}_{\tau-1} \tag{4}$$

$$\mathbf{z}_{\tau} = \mathbf{h}\left(\mathbf{x}_{\tau}\right) + \mathbf{v}_{\tau} \tag{5}$$

where $\mathbf{x}_{\tau} \in \mathbb{R}^{N}, \mathbf{z}_{\tau} \in \mathbb{R}^{M}$ are the *N*-dimensional state variables and *M*-dimensional measurement variables at $\tau-$th m oment, respectively. $\mathbf{f}(\bullet)$ and $\mathbf{h}(\bullet)$ represent the state transfer and observation equations, respectively. $\mathbf{w}_{\tau-1}$ and $\mathbf{v}_{\tau}$ represent state noise and measurement noise with covariance matrices of $\mathbf{W}_{\tau-1}$ and $\mathbf{V}_{\tau}$, respectively. $\mathbf{w}_{\tau-1}$ and $\mathbf{v}_{\tau}$ satisfy $E[\mathbf{w}_k \mathbf{v}_{\tau}] = 0$ for any $\tau$ and *k*. The main steps of the SR-GCI-IUKF are as follows [16]

1) Predict

$$\mathbf{X}^{\varsigma}_{\tau-1|\tau-1} = \begin{cases} \mathbf{x}_{\tau-1|\tau-1} & \varsigma = 0 \\ \mathbf{x}_{\tau-1|\tau-1} + \left(\sqrt{(N+\theta)}\mathbf{S}_{\tau-1|\tau-1}\right)_{\varsigma} & \varsigma = 1,\ldots,N \\ \mathbf{x}_{\tau-1|\tau-1} + \left(\sqrt{(N+\theta)}\mathbf{S}_{\tau-1|\tau-1}\right)_{\varsigma-n} & \varsigma = N+1,\ldots,2N \end{cases} \tag{6}$$

where $\mathbf{S}_{\tau-1|\tau-1}$ is obtained by Cholesky for $\mathbf{P}_{\tau|\tau-1}$, $\mathbf{P}_{\tau-1|\tau-1}$ represents the state covariance matrix. $(\bullet)_{\varsigma}$ is the $\varsigma-$th element of the matrix and $\theta = \rho^2(\eta+\kappa)-\eta$ is the scale correction factor.

$$\mathbf{x}_{\tau|\tau-1} = \sum_{\alpha=0}^{2N} \xi^{\alpha}_{m} \mathbf{f}\left(\chi^{\varsigma}_{\tau-1|\tau-1}\right) \tag{7}$$

$$\mathbf{S}_{\tau|\tau-1} = qr\left\{\left[\sqrt{\omega^{1}_{c}}\left(\mathbf{X}_{1:2N,\tau|\tau-1} - \bar{\mathbf{x}}_{\tau}\right)\right] \quad \sqrt{\mathbf{W}_{\tau-1}}\right\} \tag{8}$$

$$\mathbf{S}_{\tau|\tau-1} = cholupdate\left\{\mathbf{S}_{\tau|\tau-1} \quad \mathbf{X}_{o,\tau} - \bar{\mathbf{x}}_{\tau} \quad \omega^{0}_{c}\right\} \tag{9}$$

where $\xi^{0}_{m} = \dfrac{\lambda}{N+\lambda}$, $\xi^{0}_{c} = \dfrac{\lambda}{N+\lambda} + (1-a^2+b)$ and $\xi^{\varsigma}_{m} = \xi^{\varsigma}_{c} = \dfrac{1}{2(N+\lambda)}, \varsigma = 1 \sim 2N$.

2) Update

Similarly, $2N+1$ sigma points was obtained in neighborhood of a $\mathbf{x}_{\tau|\tau-1}$ using the UT transform.

$$\mathbf{X}^{\varsigma}_{\tau|\tau-1} = \begin{cases} \mathbf{x}_{\tau|\tau-1} & \varsigma = 0 \\ \mathbf{x}_{\tau|\tau-1} + \left(\sqrt{(N+\theta)}\mathbf{S}_{\tau|\tau-1}\right)_{\varsigma} & \varsigma = 1,\ldots,N \\ \mathbf{x}_{\tau|\tau-1} + \left(\sqrt{(N+\theta)}\mathbf{S}_{\tau|\tau-1}\right)_{\varsigma-N} & \varsigma = N+1,\ldots,2N \end{cases} \tag{10}$$

Algorithm: SR-GCI-IUKF

**1: Input:** $\mathbf{f}(\bullet)$, $\mathbf{h}(\bullet)$, $\mathbf{W}_\tau$, $\mathbf{V}_\tau$, $\varepsilon$, $\delta$, $\theta$.

**2: Output:** $\mathbf{x}_{\tau|\tau}$ for $\tau = 1,2,3,\ldots,N$.

**3: Intitialization:** Setting initial filter state values $\mathbf{x}_{0|0}$ and initial covariance matrix values $\Sigma_{0|0}$.

**4: for** $\tau = 1,2,3,\ldots,N$ **do**

The priori state estimates $\mathbf{x}_{\tau|\tau-1}$ as well as the corresponding covariance matrix $\Sigma_{\tau|\tau-1}$ are computed by (6)-(15).

**while** $\dfrac{\left\| \overset{i+1}{\mathbf{x}_\tau} - \overset{i}{\mathbf{x}_\tau} \right\|}{\left\| \overset{i}{\mathbf{x}_\tau} \right\|} > \varepsilon$ **do**

Compute the $\overset{i+1}{\mathbf{x}_\tau}$ via (32).

**end while**

Updating the covariance matrix by (34)-(35).

**end for**

Then further calculate the priori measurements $\mathbf{z}_{\tau|\tau-1}$ and the state-measurement cross-covariance matrix $\mathbf{P}_{\mathbf{xz},\tau}$ by

$$\mathbf{z}_{\tau|\tau-1} = h\left(\mathbf{X}_{\tau|\ \tau-1}\right) \tag{11}$$

$$\hat{\mathbf{z}}_{\tau|\tau-1} = \sum_{\delta=0}^{2N} \xi_m^{\varsigma} h\left(\mathbf{X}_{\tau|\ \tau-1}^{\varsigma}\right) \tag{12}$$

$$\mathbf{P}_{\mathbf{xz},\tau|\tau-1} = \sum_{\alpha=0}^{2N} \xi_c^{\varsigma} \left[\mathbf{X}_{\tau|\ \tau-1}^{\varsigma} - \mathbf{x}_{\tau|\ \tau-1}\right]\left[h\left(\mathbf{X}_{\tau|\ \tau-1}^{\varsigma}\right) - \hat{\mathbf{z}}_{\tau|\ \tau-1}\right]^T \tag{13}$$

$$\mathbf{D}_{\tau|\tau-1} = qr\left\{\left[\sqrt{\xi_c^1}\left(\mathbf{z}_{1:2N,\tau|\tau-1} - \hat{\mathbf{z}}_\tau\right)\right] \quad \sqrt{\mathbf{V}_{\tau-1}}\right\} \tag{14}$$

$$\mathbf{D}_{\tau|\tau-1} = cholupdate\left\{\mathbf{D}_{\tau|\tau-1} \quad \mathbf{X}_{o,\tau} - \bar{\mathbf{x}_\tau} \quad \omega_c^0\right\} \tag{15}$$

where $\mathbf{D}_{\tau|\tau-1} = \sqrt{\mathbf{P}_{zz,\tau|\tau-1}}$.

A nonlinear argument equation that unifies the state error and the measurement error is constructed. Consider first the nonlinear generalized model [18]

$$\begin{bmatrix} \mathbf{x}_{\tau|\tau-1} \\ \mathbf{z}_\tau \end{bmatrix} = \begin{bmatrix} \mathbf{x}_\tau \\ \mathbf{h}(\mathbf{x}_\tau) \end{bmatrix} + \begin{bmatrix} -\left(\mathbf{x}_\tau - \mathbf{x}_{\tau|\tau-1}\right) \\ \mathbf{v}_\tau \end{bmatrix} \tag{16}$$

The error covariance of $\left[-\left(\mathbf{x}_\tau - \mathbf{x}_{\tau|\tau-1}\right), \quad \mathbf{v}_\tau\right]^T$ is

$$E\left[\begin{bmatrix} -\left(\mathbf{x}_\tau - \mathbf{x}_{\tau|\tau-1}\right) \\ \mathbf{v}_\tau \end{bmatrix}\left[-\left(\mathbf{x}_\tau - \mathbf{x}_{\tau|\tau-1}\right), \quad \mathbf{v}_\tau\right]\right] = \begin{bmatrix} \mathbf{P}_{\tau|\tau-1} & 0 \\ 0 & \mathbf{V}_\tau \end{bmatrix} = \Psi_\tau \Psi_\tau^T \tag{17}$$

$$\Psi_\tau = \begin{bmatrix} \mathbf{S}_{\tau|\tau-1} & 0 \\ 0 & \Psi_{\tau|\tau-1}^{\mathbf{v}} \end{bmatrix} \tag{18}$$

where $\mathbf{S}_{\tau|\tau-1}$ and $\Psi_{\tau|\tau-1}^{\mathbf{v}}$ were obtained by Cholesky for $\mathbf{P}_{\tau|\tau-1}$ and $\mathbf{V}_\tau$ respectively.

Then multiply (16) left by $\Psi_\tau^{-1}$. We get

$$\mathbf{Q}_\tau = g_\tau(\mathbf{x}_\tau) + e_\tau \tag{19}$$

where

$$\mathbf{Q}_\tau = \left[\left(\mathbf{S}_{\tau|\tau-1}\right)^{-1} \mathbf{x}_{\tau|\tau-1}, \quad \left(\Psi_{\tau|\tau-1}^{\mathbf{v}}\right)^{-1} \mathbf{z}_\tau\right]^T \tag{20}$$

$$g_\tau(\mathbf{x}_\tau) = \left[\left(\mathbf{S}_{\tau|\tau-1}\right)^{-1} \mathbf{x}_\tau, \quad \left(\Psi_{\tau|\tau-1}^{\mathbf{v}}\right)^{-1} \mathbf{h}(\mathbf{x}_\tau)\right]^T \tag{21}$$

$$e_\tau = \Psi_\tau^{-1}\left[-\left(\mathbf{x}_\tau - \mathbf{x}_{\tau|\tau-1}\right), \quad \mathbf{v}_\tau\right]^T \tag{22}$$

We have $E\left[e_\tau e_\tau^T\right] = \mathbf{I}_{N+M}$. Therefore, $e_\tau$ is white noise, making (19) a nonlinear regression equation.

Then, we give a derivation of the state $\mathbf{x}_{\tau|\tau}$ and covariance matrix $\mathbf{P}_{\tau|\tau}$ update for SR-GCI-IUKF. Combining the above nonlinear framework with (3), the cost function is obtained

$$J_{GCIC}(\mathbf{x}_\tau) = \mu\left(1 - \sum_{i=1}^{n+m} \exp\left(-\varphi\left|e_{\tau,i}\right|^{\delta}\right)\right) \tag{23}$$

where $e_{\tau,i}$ is the *i*-th element of $e_\tau$. The optimal value for state $\mathbf{x}_{\tau|\tau}$ is obtained by

$$\mathbf{x}_{\tau|\tau} = \arg\min J_{GCIC}(\mathbf{x}_\tau) = \arg\min \mu\left(1 - \sum_{i=1}^{n+m} \exp\left(-\varphi\left|e_{\tau,i}\right|^{\delta}\right)\right) \tag{24}$$

The derivative of the cost function with respect to $x_\tau$ gives

$$\frac{\partial J_{GCIC}}{\partial \mathbf{x}_\tau} = \mu\delta\varphi \sum_{i=1}^{N+M} \exp\left(-\varphi\left|e_{\tau,i}\right|^{\delta}\right)\varphi\left|e_{\tau,i}\right|^{\delta-1} \frac{\partial e_{\tau,i}}{\partial \mathbf{x}_\tau} \tag{25}$$

Consider the following diagonal matrix

$$\Theta_{\mathbf{x},\tau} = diag\left(\phi(e_{\tau,1}), \phi(e_{\tau,2}), \cdots, \phi(e_{\tau,N})\right) \tag{26}$$

$$\Theta_{\mathbf{z},\tau} = diag\left(\phi(e_{\tau,N+1}), \phi(e_{\tau,N+2}), \cdots, \phi(e_{\tau,N+M})\right) \tag{27}$$

with

$$\Theta_\tau = \begin{bmatrix} \Theta_{\mathbf{x},\tau} & 0 \\ 0 & \Theta_{\mathbf{z},\tau} \end{bmatrix} \tag{28}$$

where

$$\phi(e_{\tau,i}) = \mu\delta\varphi\left|e_{\tau,i}\right|^{\delta-2} \exp(-\varphi\left|e_{\tau,i}\right|^{\delta}) \tag{29}$$

It shows that $S_\tau$ is independent of $x_\tau$, and we have

$$\Omega_\tau = \frac{\partial e_{\tau,i}}{\partial \mathbf{x}_\tau} = -\frac{\partial g_\tau(\mathbf{x}_\tau)}{\partial \mathbf{x}_\tau} \tag{30}$$

Making the derivative of the cost function equal to 0, the optimal value of $\mathbf{x}_{\tau|\tau}$ is obtain. Deriving a solution for $\partial J_{GCIC}/\partial \mathbf{x}_\tau = 0$ based on iterative least squares.

$$\begin{aligned} \mathbf{x}_\tau^{t+1} = \mathbf{x}_\tau^t - &\left(\left(\Omega_\tau^t\right)^T \Theta_\tau^t \Omega_\tau^t\right)^{-1} \\ &\times \left(\Omega_\tau^t\right)^T \Theta_\tau^t \left(\mathbf{S}_\tau - g_\tau\left(\mathbf{x}_\tau^t\right)\right) \end{aligned} \tag{31}$$

where *t* is the number of iterations. (31) can be rewritten by matrix inversion lemma [28] as

$$\overset{t+1}{\mathbf{x}_\tau} = \mathbf{x}_{\tau|\tau-1} + \mathbf{K}_\tau\left(\mathbf{y}_\tau^t - \mathbf{h}\left(\mathbf{x}_\tau^t\right) - \mathbf{H}\left(\mathbf{x}_\tau^t\right)\left(\mathbf{x}_{\tau|\tau-1} - \mathbf{x}_\tau^t\right)\right) \tag{32}$$

$$\mathbf{K}_\tau^t = \Sigma_{\tau|\tau-1}^t \mathbf{H}^T\left(\mathbf{x}_\tau^t\right)\left(\mathbf{H}\left(\mathbf{x}_\tau^t\right)\Sigma_{\tau|\tau-1}^t \mathbf{H}^T(\mathbf{x}_\tau^t) + \mathbf{V}_\tau^t\right)^{-1} \tag{33}$$

where $\mathbf{\Sigma}^t_{\tau|\tau-1} = \mathbf{S}_{\tau|\tau-1}\left(\Theta^t_{\mathbf{x},\tau}\right)^{-1}\left(\mathbf{S}_{\tau|\tau-1}\right)^T$, $\mathbf{V}^t_\tau = \Psi^v_{\tau|\tau-1}\left(\Theta^t_{\mathbf{z},\tau}\right)^{-1}\left(\Psi^v_{\tau|\tau-1}\right)^T$, $\mathbf{H}\left(\mathbf{x}^t_\tau\right) = \partial\mathbf{h}/\partial\mathbf{x}_\tau\,|_{\mathbf{x}_\tau=\mathbf{x}^t_\tau}$.

After an iterative step, the optimal state value is obtained and finally the covariance matrix is updated by

$$\mathbf{S}_{\tau|\tau} = cholupdate\left\{\mathbf{D}_{\tau|\tau-1} \quad \mathbf{G}_\tau \quad -1\right\} \tag{34}$$

where

$$\mathbf{G}_\tau = \mathbf{K}_\tau \mathbf{D}_{\tau|\tau-1} \tag{35}$$

**Remark III.1:** Since GCI will degenerate into CI as $\delta = 2$, our proposed algorithm will also degenerate into SR-CI-IUKF algorithm. As an enhanced version of SR-CI-IUKF, different values for $\delta$ would allow SR-GCI-IUKF algorithm to adapt to more complex noise environments.

**Remark III.2:** The SR-GCI-IUKF algorithm proposed herein addresses state estimation errors by minimizing the GCI cost function. As a result, SR-GCI-IUKF utilize the GCI to capture the information of higher order moments in the error. Unlike traditional Correntropy-based square-root KFs, SR-GCI-IUKF is insensitive to kernel bandwidth choice, thereby achieving a better combination of stability and accuracy. In addition, the constructed nonlinear error generalization model effectively improves the accuracy of the algorithm to face the nonlinear state estimation problem.

*B. Parameter optimization*

Compared with CI, GCI offers greater flexibility in handling various complex non-Gaussian noise environments. However, this advantage comes at the cost of increased parameter burden, as the algorithm requires manual parameter adjustment tailored to different noise conditions—a process that can incur significant labor costs. To address this issue, this section introduces an automated parameter adjustment method that enables the SR-GCI-IUKF to adaptively configure GCI parameters according to the prevailing noise environment.

In GCI, the information potential is given by

$$\mathrm{M}\left(\delta,\theta\right) = E\left[q_{\delta,\theta}\left(\mathbf{e}\right)\right] = \int_{-\infty}^{\infty} q_{\delta,\theta}\left(\varepsilon\right) p_e\left(\varepsilon\right) d\varepsilon \tag{36}$$

where $q_{\delta,\theta}\left(e\right) = \frac{\mu}{N}\left(1 - \sum_{i=1}^{N}\exp\left(-\varphi\left|e_i\right|^\delta\right)\right)$, $p_e\left(\varepsilon\right)$ is the error probability density function (PDF). Then, using the idea of information potential energy matching, the parameter optimization objective is redefined as minimizing the distance between the GCI estimation and the real error PDF:

$$T_{\delta,\theta}\left(X,Y\right) = \frac{1}{2}\int_{-\infty}^{\infty}\left[p_e\left(\varepsilon\right)\right]^2 d\varepsilon - \frac{1}{2}\int_{-\infty}^{\infty}\left[q_{\delta,\theta}\left(\varepsilon\right) - p_e\left(\varepsilon\right)\right]^2 d\varepsilon \tag{37}$$

Since $p_e\left(\varepsilon\right)$ is parameter independent, maximizing $T_{\delta,\theta}\left(X,Y\right)$ is equivalent to

$$\begin{aligned} \arg\max\left\{T_{\delta,\theta}\left(X,Y\right)\right\} &= \arg\max\left\{\frac{1}{2}\int_{-\infty}^{\infty}\left[q_{\delta,\theta}\left(\varepsilon\right) - p_e\left(\varepsilon\right)\right]^2 d\varepsilon\right\} \\ &= \arg\max\left\{\int_{-\infty}^{\infty} q_{\delta,\theta}\left(\varepsilon\right) p_e\left(\varepsilon\right) d\varepsilon - \frac{1}{2}\int_{-\infty}^{\infty}\left[q_{\delta,\theta}\left(\varepsilon\right)\right]^2 d\varepsilon\right\} \end{aligned} \tag{38}$$

Given $N$-th error samples $\left\{e_i\right\}_{i=1}^N$, we can estimate the PDF of the error as

$$p_e\left(\varepsilon\right) = \frac{1}{N}\sum_{i=1}^{N}\rho\left(\varepsilon - e_i\right) \tag{39}$$

where $\rho\left(\bullet\right)$ is the Dirac delta function, which represents the GCI estimation as

$$Q_{\delta,\theta}\left(\varepsilon\right) = \frac{1}{N}\sum_{i=1}^{N} q_{\delta,\theta}\left(e_i\right) \tag{40}$$

Express the quadratic term estimation as

$$V_{\delta,\theta} = \frac{1}{N^2}\sum_{i=1}^{N}\sum_{j=1}^{N} q_{\delta,\theta}\left(e_i - e_j\right) \tag{41}$$

Therefore, we approximate the optimization problem as

$$\arg\max_{\delta,\theta} T_{\delta,\theta}\left(X,Y\right) = Q_{\delta,\theta}\left(X,Y\right) - \frac{1}{2}V_{\delta,\theta} \tag{42}$$

Therefore, we realize the optimization of parameter $\delta,\theta$ and adopt the alternative optimization strategy. First, fix $\delta$, we need to search for the optimal $\theta$ on the finite set $S_\theta$

$$\hat{\theta} = \arg\max_{\theta\in S_\theta} T_{\delta,\theta}\left(X,Y\right) \tag{43}$$

Then fix $\theta$ and search for the optimal $\delta$ on the finite set $S_\delta$

$$\hat{\delta} = \arg\max_{\delta\in S_\delta} T_{\delta,\theta}\left(X,Y\right) \tag{44}$$

So far, the adaptive optimal selection of parameters is realized.

The parameter adaptation strategy employed in SR-GCI-IUKF (optimization) searches for the optimal shape parameter $\delta$ and scale parameter $\theta$ within predefined candidate sets. In practice, the appropriate ranges for these parameters depend on the noise characteristics of the specific application. As a general guideline, $\delta$ can be selected from $S_\delta$ such as (1, 3), which covers kernel shapes from Laplacian-like to nearly Gaussian. The scale parameter $\theta$, which controls the kernel width, may be chosen from $S_\theta$ such as (5, 25) after normalizing the innovation error.

Parameter optimization is performed once per filtering iteration, enabling the filter to adapt quickly to time-varying noise conditions. However, this repeated optimization over the candidate sets contributes directly to the increased computational cost observed in Table VII. In real-time applications where computational resources are limited, the search ranges can be narrowed based on prior knowledge of the noise environment, or the optimization can be executed only when a significant change in the innovation statistics is detected, thereby reducing the average runtime while retaining most of the robustness benefits.

*C. Performance Analysis*

This subsection assesses the convergence properties of SR-GCI-IUKF.

In order to analyze the convergence of algorithm errors, we first define two forms of error, the priori estimation error as well as the posteriori estimation error

$$\chi_{\tau|\tau-1} = \mathbf{x}_\tau - \hat{\mathbf{x}}_{\tau|\tau-1} \tag{45}$$

$$\chi_{\tau|\tau} = \mathbf{x}_\tau - \hat{\mathbf{x}}_{\tau|\tau} \tag{46}$$

We can easily obtain the relationship between the two errors

$$\chi_{\tau|\tau}=\left(\mathrm{I}_N-\mathbf{K}_\tau\mathbf{H}_\tau\right)\chi_{\tau|\tau-1}-\mathbf{K}_\tau\mathrm{v}_\tau \tag{47}$$

where $\mathbf{H}_\tau=\partial\mathbf{h}/\partial\mathbf{x}_\tau\,|_{\mathbf{x}_\tau=\mathbf{x}_{\tau|\tau-1}}$. Here we give the following definition.

***Definition 1:*** $l_\tau$ is considered to be exponentially bounded in mean square, if $E\left(\|l_0\|^2\right)<\infty$, and there are $\eta,\iota>0$ and $0<\nu<1$, The following equation holds

$$E\left(\|l_\tau\|^2\right)\le\eta E\left(\|l_o\|^2\right)\nu^\tau+\iota \tag{48}$$

This definition aims to mathematically ensure that the estimation error does not grow unbounded, but rather decays exponentially over time or stabilizes within a certain boundary, providing a strict criterion for "filter stability".

The subsequent derivation is based on the following assumptions, which aim to constrain the physical feasibility of the system model and noise, ensure that the state transition, observation function, and noise covariance matrix change within a reasonable range, thereby ensuring that the error covariance matrix remains well defined and avoids dispersion during iteration.

Then we consider some assumption.

***Assumption 1:*** Considering some positive real numbers: $\underline{\mathrm{f}},\bar{\mathrm{f}},\underline{\mathrm{h}},\bar{\mathrm{h}},\underline{\mathrm{p}},\bar{\mathrm{p}},\underline{\mathrm{w}},\bar{\mathrm{w}},\underline{\mathrm{v}},\bar{\mathrm{v}},\lambda_{\mathrm{w}},\lambda_{\mathrm{v}}>0$ .For every $\tau>0$, the above matrix satisfies [24][25][26]

$$\underline{\mathrm{f}}\le\left\|\mathbf{F}_\tau\right\|\le\bar{\mathrm{f}},\underline{\mathrm{h}}\le\left\|\mathbf{H}_\tau\right\|\le\bar{\mathrm{h}} \tag{49}$$

$$\underline{\mathrm{p}}\,\mathbf{I}_N\le\mathbf{P}_{\tau|\tau-1}\le\bar{\mathrm{p}}\,\mathbf{I}_N \tag{50}$$

$$\underline{\mathrm{w}}\,\mathbf{I}_N\le\mathbf{W}_\tau\le\bar{\mathrm{w}}\,\mathbf{I}_N \tag{51}$$

$$\underline{\mathrm{v}}\,\mathbf{I}_M\le\mathbf{V}_\tau\le\bar{\mathrm{v}}\,\mathbf{I}_M \tag{52}$$

$$E\left[\mathbf{v}_\tau^T\mathbf{v}_\tau\right]\le\lambda_{\mathbf{v}},E\left[\mathbf{w}_\tau^T\mathbf{w}_\tau\right]\le\lambda_{\mathbf{w}} \tag{53}$$

$$\underline{\vartheta}_{\mathbf{x}}\le\left\|\Theta_{\mathbf{x},\tau}\right\|\le\bar{\vartheta}_{\mathbf{x}},\underline{\vartheta}_{\mathbf{z}}\le\left\|\Theta_{\mathbf{z},\tau}\right\|\le\bar{\vartheta}_{\mathbf{z}} \tag{54}$$

where $\mathbf{F}_{\tau-1}=\partial\mathbf{f}/\partial\mathbf{x}_\tau\,|_{\mathbf{x}_\tau=\mathbf{x}_{\tau-1|\tau-1}}$.

***Assumption 2:*** For every $\tau>0$, $\mathbf{F}_\tau$ is non-singular.

**Remark III.3:** Since this energy constraint is usually present in real systems, the assumption of boundedness for the noise covariance matrix usually holds. Moreover, the system state update equations and the measurement equations are usually canonical, so the assumption of boundedness is also reasonable.

Next, we analyze the boundedness of the Kalman gain $\mathbf{K}_\tau$. According to (33), we can see that the need to prove that the $\mathbf{K}_\tau$ is bounded necessitates an analysis of $\mathbf{\Sigma}_{\tau|\tau-1}$ and $\mathbf{V}_\tau$. It is easy to know $\mathbf{\Sigma}_{\tau|\tau-1}$ and $\mathbf{V}_\tau$ satisfy

$$\frac{\underline{\mathrm{p}}}{\bar{\vartheta}_{\mathbf{x}}}\le\left\|\mathbf{\Sigma}_{\tau|\tau-1}\right\|\le\frac{\bar{\mathrm{p}}}{\underline{\vartheta}_{\mathbf{x}}} \tag{55}$$

$$\frac{\underline{\mathrm{v}}}{\bar{\vartheta}_{\mathbf{z}}}\le\left\|\mathbf{V}_\tau\right\|\le\frac{\bar{\mathrm{v}}}{\underline{\vartheta}_{\mathbf{z}}} \tag{56}$$

So according to (33), we have

$$\frac{\underline{\mathrm{p}}\underline{\mathrm{h}}}{\bar{\mathrm{h}}^2\bar{\mathrm{p}}+\frac{\bar{\vartheta}_{\mathbf{x}}}{\underline{\vartheta}_{\mathbf{z}}}\bar{\mathrm{v}}}\le\mathbf{K}_\tau\le\frac{\bar{\mathrm{p}}\bar{\mathrm{h}}}{\underline{\mathrm{h}}^2\underline{\mathrm{p}}+\frac{\underline{\vartheta}_{\mathbf{x}}}{\bar{\vartheta}_{\mathbf{x}}}\underline{\mathrm{v}}} \tag{57}$$

For the purpose of subsequent calculations, we define

$$\underline{\lambda}_{\mathbf{K}_\tau}=\frac{\underline{\mathrm{p}}\underline{\mathrm{h}}}{\bar{\mathrm{h}}^2\bar{\mathrm{p}}+\frac{\bar{\vartheta}_{\mathbf{x}}}{\underline{\vartheta}_{\mathbf{z}}}\bar{\mathrm{v}}} \tag{58}$$

$$\bar{\lambda}_{\mathbf{K}_\tau}=\frac{\bar{\mathrm{p}}\bar{\mathrm{h}}}{\underline{\mathrm{h}}^2\underline{\mathrm{p}}+\frac{\underline{\vartheta}_{\mathbf{x}}}{\bar{\vartheta}_{\mathbf{x}}}\underline{\mathrm{v}}} \tag{59}$$

Consequently, we can conclude that the Kalman gain $\mathbf{K}_{\tau|\tau-1}^\iota$ is bounded.

***Lemma 1:*** Considering a Lyapunov function $\hbar_\tau\left(\mathbf{x}_\tau\right)$, and a real number $\underline{o},\bar{o}$ for every $\tau>0$

$$\underline{o}\left\|\mathbf{x}_\tau\right\|^2\le\hbar_\tau\left(\mathbf{x}_\tau\right)\le\bar{o}\left\|\mathbf{x}_\tau\right\|^2 \tag{60}$$

And there exists a corresponding number $\iota_\tau>0$ and $1>\gamma_\tau>0$. We get

$$E\left[\hbar_{\tau+1}\left(\mathbf{x}_{\tau+1}\right)|\,\mathbf{x}_\tau\right]\le\left(1-\gamma_\tau\right)\hbar_\tau\left(\mathbf{x}_\tau\right)+\iota_\tau \tag{61}$$

Then ask for the expectation of obtaining

$$E\left[\hbar_{\tau+1}\left(\mathbf{x}_{\tau+1}\right)\right]\le\left(1-\gamma_\tau\right)E\left[\hbar_\tau\left(\mathbf{x}_\tau\right)\right]+\iota_\tau \tag{62}$$

Iterating over (51) yields

$$E\left[\hbar_{\tau+1}\left(\mathbf{x}_{\tau+1}\right)\right]\le E\left[\hbar_0\left(\mathbf{x}_0\right)\right]\prod_{k=0}^{\tau}\left(1-\gamma_k\right)+\sum_{k=0}^{\tau}\iota_k\left(1-\gamma_k\right)^k \tag{63}$$

According to (49), we have

$$\underline{o}E\left[\left\|\mathbf{x}_\tau\right\|^2\right]\le\bar{o}E\left[\left\|\mathbf{x}_0\right\|^2\right]\prod_{k=0}^{\tau}\left(1-\gamma_k\right)+\sum_{k=0}^{\tau}\iota_k\left(1-\gamma_k\right)^k \tag{64}$$

As a result, we get

$$E\left[\left\|\mathbf{x}_\tau\right\|^2\right]\le\frac{\bar{o}}{\underline{o}}E\left[\left\|\mathbf{x}_0\right\|^2\right]\left(1-\gamma\right)^\tau+\frac{\iota}{\underline{o}\gamma} \tag{65}$$

where $\gamma=\min\left(\gamma_0,\gamma_1,\ldots,\gamma_\tau\right)$, $\iota=\max\left(\iota_0,\iota_1,\ldots,\iota_\tau\right)$.

***Lemma 2:*** By Assumptions 1 and 2, for any integer $\tau\ge0$, there exists $1>\gamma_\tau>0$ such that $\mathbf{P}_{\tau+1|\tau}^{-1}$ meet

$$\mathrm{M}_\tau^T\mathbf{P}_{\tau+1|\tau}^{-1}\mathrm{M}_\tau\le\left(1-\gamma_\tau\right)\mathbf{P}_{\tau|\tau-1}^{-1} \tag{66}$$

where $\mathrm{M}_\tau=\mathbf{F}_{\tau+1}\left(\mathbf{I}_N-\mathbf{K}_\tau\mathbf{H}_\tau\right)$, $\gamma_\tau=\dfrac{\underline{\mathrm{w}}}{\bar{\mathrm{p}}\left(\bar{\mathrm{f}}+\bar{\mathrm{f}}\bar{\lambda}_{\mathbf{K}_\tau}\bar{\mathrm{h}}\right)^2+\underline{\mathrm{w}}}$.See Appendix for the proof process.

***Lemma 3:*** By assumption 1, for any integer $\tau\ge0$

$$\begin{aligned}&E\left[\left(\mathbf{F}_{\tau+1}\mathbf{K}_\tau\mathbf{v}_\tau\right)^T\mathbf{P}_{\tau+1|\tau}^{-1}\left(\mathbf{F}_{\tau+1}\mathbf{K}_\tau\mathbf{v}_\tau\right)|\,\chi_{\tau|\tau-1}\right]\\&=E\left[\mathbf{v}_\tau^T\left(\mathbf{F}_{\tau+1}\mathbf{K}_\tau\right)^T\mathbf{P}_{\tau+1|\tau}^{-1}\left(\mathbf{F}_{\tau+1}\mathbf{K}_\tau\right)\mathbf{v}_\tau\right]\\&\le\frac{\left(\bar{\mathrm{f}}\bar{\lambda}_{\mathbf{K}_\tau}\right)^2}{\underline{\mathrm{p}}}\lambda_{\mathbf{v}}\end{aligned} \tag{67}$$

We then proceed to prove the convergence of the errors. According to Assumption 1 and Assumption 2, if the priori state error satisfies $E\left[\left\|\chi_{1|0}\right\|^2\right]<\infty$, we get

$$E\left[\left\|\chi_{\tau+1|\tau}\right\|^2\right]\le\frac{\bar{p}}{\underline{p}}\left(1-\gamma_\tau\right)E\left[\left\|\chi_{\tau|\tau-1}\right\|^2\right]+\bar{p}\iota_\tau \tag{68}$$

where $\iota_\tau=\dfrac{\left(\bar{f}\bar{\lambda}_{\mathbf{K}_\tau}\right)^2\lambda_{\mathbf{v}}+\lambda_{\mathbf{w}}}{\underline{p}}$. The posteriori state estimation error satisfies

$$E\left[\left\|\chi_{\tau|\tau}\right\|^2\right]\le 2\underline{f}^{-2}\Upsilon_\tau \tag{69}$$

where $\Upsilon_\tau=\dfrac{\bar{p}}{\underline{p}}E\left[\left\|\chi_{1|0}\right\|^2\right]\prod_{k=0}^{\tau-1}\left(1-\gamma_k\right)+\left(\bar{p}\sum_{k=0}^{\tau-1}\iota_k\left(1-\gamma_k\right)+\lambda_{\mathbf{w}}\right)$.

Therefore, if the error weights $\dfrac{\bar{\vartheta}_{\mathbf{x}}}{\underline{\vartheta}_{\mathbf{z}}}$ are bounded, $\chi_{\tau|\tau}$ is exponentially bounded on the mean square, i.e.,

$$E\left[\left\|\chi_{\tau|\tau}\right\|^2\right]\le 2\underline{f}^{-2}\Lambda_\tau \tag{70}$$

where $\Lambda_\tau=\dfrac{\bar{p}}{\underline{p}}E\left[\left\|\chi_{1|0}\right\|^2\right]\prod_{k=0}^{\tau-1}\left(1-\gamma_k\right)+\left(\bar{p}\dfrac{\iota}{\gamma}+\lambda_{\mathbf{w}}\right)$. We defining

$$0<\underline{s}\le\frac{\bar{\vartheta}_{\mathbf{x}}}{\underline{\vartheta}_{\mathbf{z}}}\le\bar{s} \tag{71}$$

Therefore, we have

$$\iota=\frac{\left(\dfrac{\bar{f}\bar{p}\bar{h}}{\underline{h}^2\underline{p}+s\underline{v}}\right)\lambda_v+\lambda_w}{\underline{p}} \tag{72}$$

$$\gamma=\frac{\underline{q}}{\bar{p}\left(\bar{f}+\dfrac{\bar{f}\bar{p}\bar{h}^2}{\underline{h}^2\underline{p}+s\underline{v}}\right)^2+\underline{q}} \tag{73}$$

Next we give a proof of the above conclusion. Consider first the following Lyapunov equation

$$g\left[\chi_{\tau+1|\tau}\right]=\chi^T_{\tau+1|\tau}\mathbf{P}^{-1}_{\tau+1|\tau}\chi_{\tau+1|\tau} \tag{74}$$

Based on the previous assumptions, we can get

$$\frac{\left\|\chi_{\tau+1|\tau}\right\|^2}{\bar{p}}\le g\left[\chi_{\tau+1|\tau}\right]\le\frac{\left\|\chi_{\tau+1|\tau}\right\|^2}{\underline{p}} \tag{75}$$

Next consider the conditional expectation of error

$$\begin{aligned}&E\left[g\left[\chi_{\tau+1|\tau}\right]\mid\chi_{\tau|\tau-1}\right]\\&=\chi^T_{\tau|\tau-1}\left(\mathbf{F}_{\tau+1}-L_\tau\right)^T\mathbf{P}^{-1}_{\tau+1|\tau}\left(\mathbf{F}_{\tau+1}-L_\tau\right)\chi_{\tau|\tau-1}\\&+E\left[L^T_\tau\mathbf{P}^{-1}_{\tau+1|\tau}L_\tau\mid\chi_{\tau|\tau-1}\right]+E\left[\mathbf{W}^T_\tau\mathbf{P}^{-1}_{\tau+1|\tau}\mathbf{W}_\tau\mid\chi_{\tau|\tau-1}\right]\end{aligned} \tag{76}$$

where $L_\tau=\mathbf{F}_{\tau+1}\mathbf{K}_\tau\mathbf{H}_{\tau+1}$. According to Lemma 2 and Lemma 3, we have

$$\begin{aligned}&E\left[g\left[\chi_{\tau+1|\tau}\right]\mid\chi_{\tau|\tau-1}\right]\le\left(1-\gamma_\tau\right)\mathbf{P}^{-1}_{\tau|\tau-1}\\&+\frac{\left(\bar{f}\bar{\lambda}_{\mathbf{K}_\tau}\right)^2}{\underline{p}}\lambda_{\mathbf{v}}+E\left[\mathbf{W}^T_\tau\mathbf{P}^{-1}_{\tau+1|\tau}\mathbf{W}_\tau\mid\chi_{\tau|\tau-1}\right]\\&\le\left(1-\gamma_\tau\right)\mathbf{P}^{-1}_{\tau|\tau-1}+\frac{\left(\bar{f}\bar{\lambda}_{\mathbf{K}_\tau}\right)^2}{\underline{p}}\lambda_{\mathbf{v}}+\frac{\lambda_{\mathbf{w}}}{\underline{p}}=\left(1-\gamma_\tau\right)\mathbf{P}^{-1}_{\tau|\tau-1}+\iota_\tau\end{aligned} \tag{77}$$

We then give the proof of (59). Based on (66), we can derive

$$\begin{aligned}&\frac{E\left[\left\|\chi_{\tau+1|\tau}\right\|^2\right]}{\bar{p}}\le E\left[g\left[\chi_{\tau+1|\tau}\right]\mid\chi_{\tau+1|\tau}\right]\\&\le\left(1-\gamma_\tau\right)g\left[\chi_{\tau+1|\tau}\right]\le\frac{\left(1-\gamma_\tau\right)}{\underline{p}}\left\|\chi_{\tau+1|\tau}\right\|^2\end{aligned} \tag{78}$$

$$\frac{E\left[\left\|\chi_{\tau+1|\tau}\right\|^2\right]}{\bar{p}}\le\frac{\left(1-\gamma_\tau\right)}{\underline{p}}E\left[\left\|\chi_{\tau+1|\tau}\right\|^2\right]+\iota_\tau \tag{79}$$

Then the prove of (60) is given. According to lemma 1, we can get

$$E\left[\left\|\chi_{\tau+1|\tau}\right\|^2\right]\le\frac{\bar{p}}{\underline{p}}E\left[\left\|\chi_{1|0}\right\|^2\right]\prod_{k=0}^{\tau-1}\left(1-\gamma_k\right)+\bar{p}\sum_{k=0}^{\tau-1}\iota_k\left(1-\gamma_k\right) \tag{80}$$

$$\left\|\chi_{\tau+1|\tau}\right\|=\left\|\mathbf{F}_{\tau+1}\chi_{\tau|\tau}+\mathbf{w}_\tau\right\|\ge\left\|\mathbf{F}_{\tau+1}\chi_{\tau|\tau}\right\|-\left\|\mathbf{w}_\tau\right\| \tag{81}$$

Thus, it is possible to obtain

$$\left\|\mathbf{F}_{\tau+1}\chi_{\tau|\tau}\right\|^2\le 2\left(\left\|\chi_{\tau|\tau}\right\|^2+\left\|\mathbf{w}_\tau\right\|^2\right) \tag{82}$$

Then we can easily get

$$E\left[\left\|\chi_{\tau+1|\tau}\right\|^2\right]\le 2\underline{f}^{-2}\left(E\left[\left\|\chi_{\tau+1|\tau}\right\|^2\right]+E\left[\left\|\mathbf{w}_\tau\right\|^2\right]\right) \tag{83}$$

According to (71), we have

$$E\left[\left\|\chi_{\tau|\tau}\right\|^2\right]\le 2\underline{f}^{-2}\Upsilon_\tau \tag{84}$$

At this point, by analyzing the boundedness of the mean square of the error, the stability of SR-GCI-IUKF is proved.

In summary, the performance analysis provides a theoretical guarantee that the estimation error of SR-GCI-IUKF remains bounded in mean square under reasonable system and noise assumptions. This ensures that the filter will not diverge even in the presence of non-Gaussian disturbances and strong nonlinearities, which is crucial for real-world applications where stability is mandatory. The proof leverages a Lyapunov-type argument and shows that the generalized correntropy-based update, together with the square-root formulation, preserves the numerical and stochastic stability of the filtering process. Thus, the analysis not only supports the empirical robustness observed in simulations but also offers a formal justification for deploying SR-GCI-IUKF in safety-critical or noise-varying environments.

### *D. Computational Complexity Analysis*

Although the proposed SR-GCI-IUKF enhances numerical stability and robustness, it is essential to evaluate its computational burden, especially for high-dimensional systems. The computational complexity of the algorithm is primarily determined by the following operations:

Sigma point propagation and unscen ted transform:

$$O\left((2N+1)\times\left(n_f+n_h\right)\right) \quad (85)$$

where $n_f, n_h$ are the cost of evaluating $\mathbf{f}(\bullet)$ and $\mathbf{h}(\bullet)$.

Square-root updates via QR and Cholesky factorization: $O\left(N^3\right)$ for covariance updates, which is comparable to standard SR-UKF.

GCI-based iterative correction: Each iteration requires solving a weighted least squares problem, involving matrix inversions of order $O\left((N+M)^3\right)$.

Therefore, the overall complexity of SR-GCI-IUKF relative to traditional KF is controlled by the above steps, resulting in a complexity of about

$$O\left((2N+1)\times\left(n_f+n_h\right)\right)+O\left(N^3\right)+O\left(t_{\max}\times(N+M)^3\right) \quad (86)$$

where $t_{\max}$ is typically small due to rapid convergence.

## IV. Simulation Results

In this subsection, experiments were conducted in the nonlinear system, land vehicle navigation system and forecasting-aided state estimation (FASE) of IEEE 14 bus test system to verify the performance of SR-GCI-IUKF by comparing it with UKF [3], IUKF [3], MCUKF [3], MEEUKF [9], and SR-CI-IUKF algorithms, respectively. And the SR-GCI-IUKF algorithm was evaluated in two configurations: without parameter adaptation and with parameter adaptation, record as SR-GCI-IUKF (trail) and SR-GCI-IUKF (optimization), respectively. To enhance the statistical significance and accuracy of experimental results, we performed M=100 independent Monte Carlo trials.

In this section, the algorithm performance is evaluated by the root mean square error, defined as follows

$$\mathrm{RMSE}\left(\mathbf{x}_\tau\right)=\sqrt{\frac{1}{\mathrm{M}L}\sum_{n=1}^{\mathrm{M}}\|\overset{c}{\mathbf{x}_\tau}-\mathbf{x}_\tau^c\|_2^2} \quad (87)$$

where $\overset{c}{\mathbf{x}_\tau}$ and $\mathbf{x}_\tau^c$ are the estimated and true values of the state, respectively. $c$ represents the $c$-th ind ependent experiment. An average RMSE is further defined

$$\mathrm{ARMSE}=\frac{1}{N}\sum_{\tau=1}^{N}\mathrm{RMSE}\left(\mathbf{x}_\tau\right) \quad (88)$$

where $N$ represents the total time samples.

### A. Nonlinear system

Consider the following nonlinear system

$$\mathbf{x}_\tau=0.5\mathbf{x}_{\tau-1}+\frac{25\mathbf{x}_{\tau-1}}{1+\mathbf{x}_{\tau-1}^2}+8\cos(1.2\mathbf{x}_{\tau-1})+\mathbf{w}_{\tau-1} \quad (89)$$

$$\mathbf{z}_\tau=\frac{\mathbf{x}_\tau^2}{20}+\mathbf{v}_\tau \quad (90)$$

TABLE II
NOISE TYPES IN DIFFERENT SITUATIONS

| | |
|---|---|
| Noise a | $\mathbf{v}_k \sim 0.9\times N(0,0.1)+0.1\times N(0,100)$ |
| Noise b | $\mathbf{v}_k \sim 0.8\times N(0,0.1)+0.2\times N(0,100)$ |
| Noise c | $\mathbf{v}_k \sim 0.7\times N(0,0.1)+0.3\times N(0,100)$ |
| Noise e | $\mathbf{v}_k \sim 0.9\times N(0,0.1)+0.1\times N(0,100)$ |
| Noise f | $\mathbf{v}_k \sim 0.9\times N(-0.1,0.1)+0.1\times N(0.1,100)$ |
| Noise g | $\mathbf{v}_k \sim 0.49\times N(-0.1,0.1)+0.49\times N(0.1,0.1)+0.02\times N(0,100)$ |

TABLE III
ALGORITHM PARAMETER SETTING

| Algorithm | kernel width of nonlinear system and Vehicle navigation system | kernel width of power system simulation |
|---|---|---|
| MCC-UKF | σ = 30 | σ = 15 |
| MEE-UKF | σ = 2 | σ = 2 |
| SR-CI-IUKF | σ = 5 | σ = 10 |
| SR-GCI-IUKF | $\delta=15, \theta=1.8$ | $\delta=1.8, \theta=10$ |

TABLE IV
COMPARISON OF ARMSE FOR DIFFERENT ALGORITHMS

| Algorithm | ARMSE of noise a | ARMSE of noise b | ARMSE of noise c |
|---|---|---|---|
| UKF | 13.1070 | 12.9954 | **12.8912** |
| IUKF | 14.1469 | 13.9255 | 14.0530 |
| MCC-UKF | 14.8160 | 13.7330 | 13.3181 |
| MEE-UKF | 14.6712 | 13.7396 | 13.7486 |
| SR-CI-IUKF | 14.4388 | 13.9552 | 13.8773 |
| SR-GCI-IUKF(Trail) | 13.0963 | 12.8249 | 13.2776 |
| SR-GCI-IUKF (opti) | **13.0570** | **12.7758** | **13.2042** |

TABLE V
COMPARISON OF ARMSE OF POSITION FOR DIFFERENT PARAMETERS

| | $\sigma=12$ | $\sigma=15$ | $\sigma=20$ |
|---|---|---|---|
| $\delta=1.7$ | nan | nan | 1.7885 |
| $\delta=1.8$ | 1.8595 | 1.7850 | 1.3176 |
| $\delta=1.9$ | 1.2261 | 1.1417 | 1.1033 |
| $\delta=2.0$ | 0.8847 | 0.9531 | 0.8338 |
| $\delta=2.1$ | 0.7613 | 0.7608 | 0.7971 |
| $\delta=2.2$ | 0.7190 | 0.6709 | 0.6711 |
| $\delta=2.3$ | 0.6255 | **0.6162** | 0.6456 |

TABLE VI
COMPARISON OF ARMSE OF VELOCITY FOR DIFFERENT PARAMETERS

| | $\sigma=12$ | $\sigma=15$ | $\sigma=20$ |
|---|---|---|---|
| $\delta=1.7$ | nan | nan | 1.1563 |
| $\delta=1.8$ | 1.0940 | 1.1046 | 1.0082 |
| $\delta=1.9$ | 1.0189 | 0.9659 | 0.9455 |
| $\delta=2.0$ | 0.8492 | 0.8997 | 0.8376 |
| $\delta=2.1$ | 0.8279 | 0.8311 | 0.8361 |
| $\delta=2.2$ | 0.7890 | 0.7865 | **0.7519** |
| $\delta=2.3$ | 0.7671 | 0.7762 | 0.7750 |

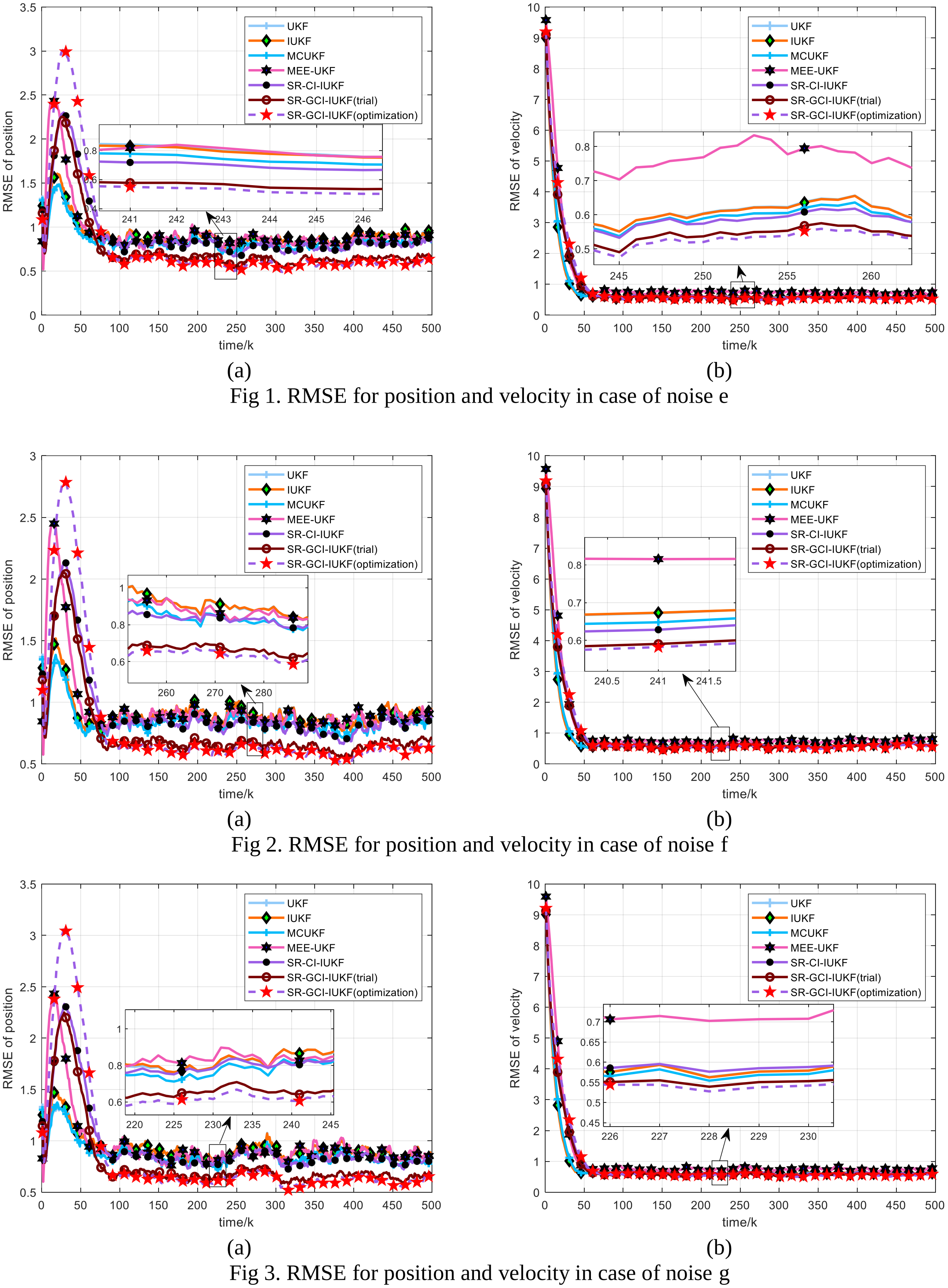

(a) (b)
Fig 1. RMSE for position and velocity in case of noise e

(a) (b)
Fig 2. RMSE for position and velocity in case of noise f

(a) (b)
Fig 3. RMSE for position and velocity in case of noise g

The model is representative of a strongly nonlinear system. In this section of the experiment, we set the initial parameters as follows: $\mathbf{x}_0 = 0$, $\mathbf{x}_{1|0} = 0$, $\mathbf{P}_{1|0} = 1$. The noise cases considered in this section are all known and the noise covariance is calculated as $\mathbf{W} = \text{var}(\mathbf{w}_\tau)$, $\mathbf{V} = \text{var}(\mathbf{v}_\tau)$. $\text{var}(\bullet)$ denotes the covariance matrix of the noise. In this subsection, the system is used to test the ability of the algorithms to cope with non-Gaussian noise, and the added non-

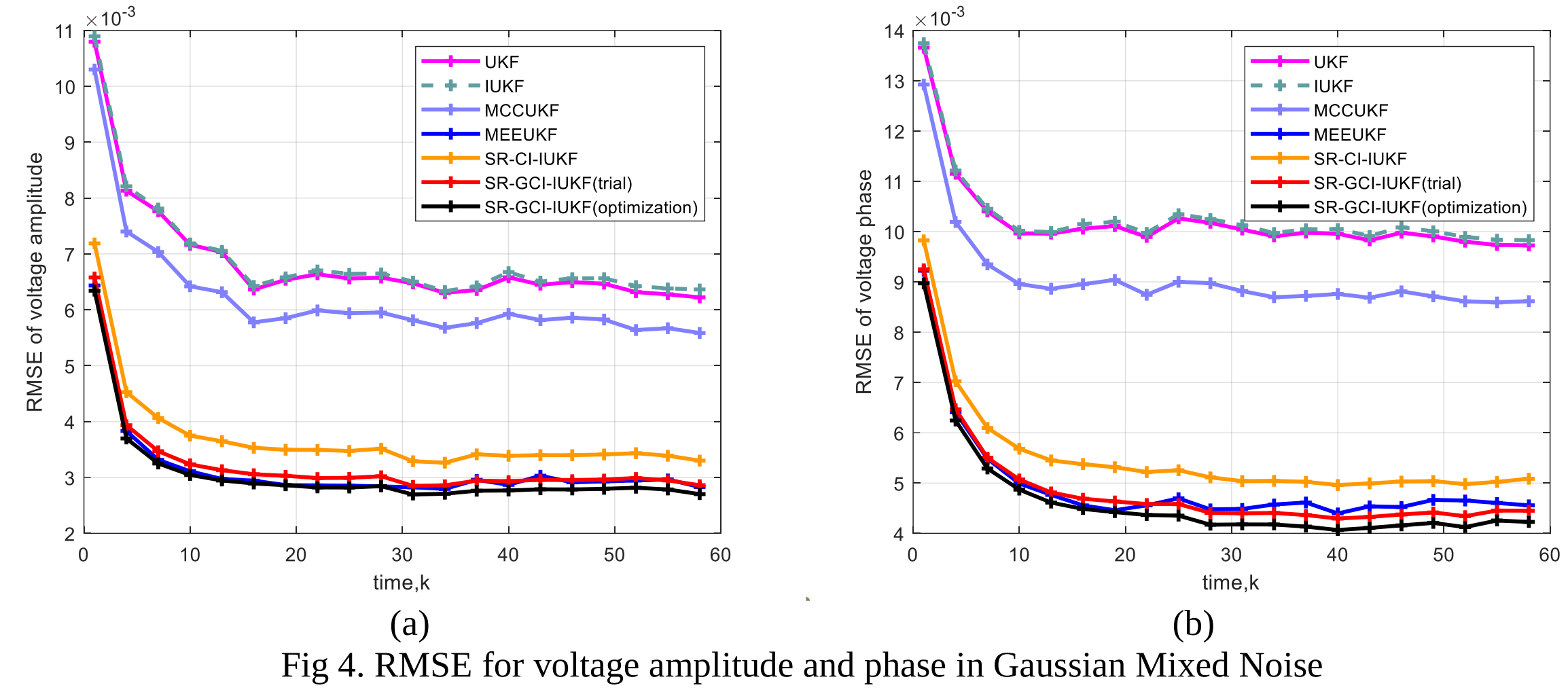

(a) (b)

Fig 4. RMSE for voltage amplitude and phase in Gaussian Mixed Noise

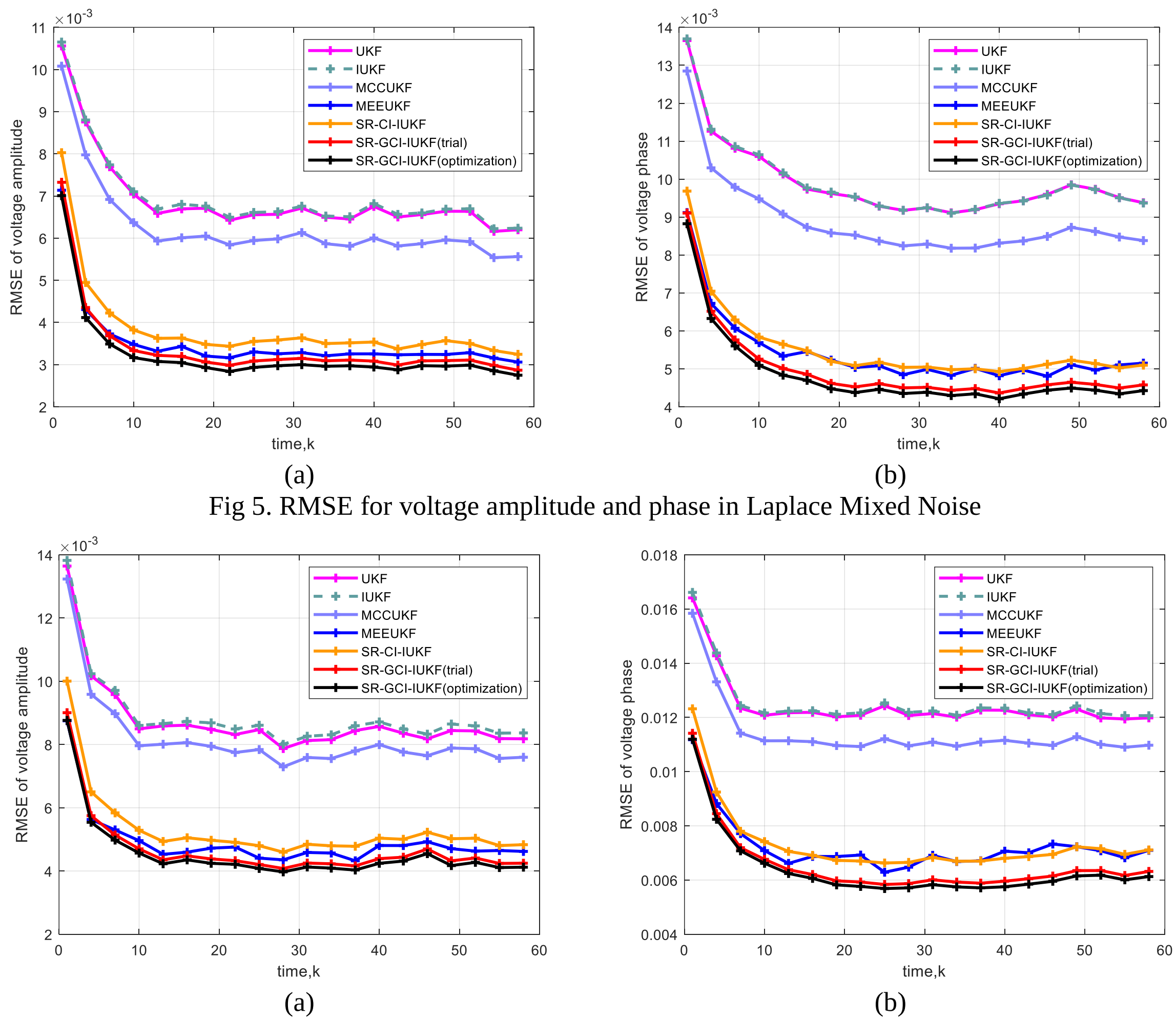

(a) (b)

Fig 5. RMSE for voltage amplitude and phase in Laplace Mixed Noise

(a) (b)

Fig 6. RMSE for voltage amplitude and phase in Asymmetric Gaussian Noise

Gaussian noise is shown in TABLE II. The kernel parameters for each algorithm are shown in TABLE III, with threshold all set to $10^{-6}$. TABLE IV then shows the ARMSE of different algorithms for different cases of non-Gaussian noise. It makes out that the SR-GCI-IUKF algorithm has a similar performance to the other robust algorithms and is slightly stronger than the other robust algorithms when confronted with non-Gaussian noise. And we can see that the parameter adaptive method can help the algorithm find the best parameters, so as to obtain the best performance.

*B. Vehicle navigation system*

The state model are as follows

$$\mathbf{x}_{\tau} = \mathbf{F}\mathbf{x}_{\tau-1} + \mathbf{w}_{\tau-1} \tag{91}$$

$$\mathbf{z}_\tau = \mathbf{h}(\mathbf{x}_\tau) + \mathbf{v}_\tau \tag{92}$$

where

$$\mathbf{F} = \begin{bmatrix} 1 & 0 & T & 0 \\ 0 & 1 & 0 & T \\ 0 & 0 & 1 & 0 \\ 0 & 0 & 0 & 1 \end{bmatrix} \tag{93}$$

a nd $\mathbf{h}(\mathbf{x}_\tau)$ is given by

$$\mathbf{h}(\mathbf{x}_\tau) = \begin{bmatrix} -\mathbf{x}_{\tau,1} - \mathbf{x}_{\tau,3} \\ -\mathbf{x}_{\tau,2} - \mathbf{x}_{\tau,4} \\ \sqrt{\mathbf{x}_{\tau,1}^2 + \mathbf{x}_{\tau,2}^2} \\ \arctan\left(\sqrt{\dfrac{\mathbf{x}_{\tau,1} + 100}{\mathbf{x}_{\tau,2} + 100}}\right) \end{bmatrix} \tag{94}$$

where $\mathbf{x}_\tau$ contains velocity and distance in the east and north directions. $T = 0.1$ is the sampling interval. The real state value, the initial state value, and the initial covariance matrix are set as $\mathbf{x}_0 = \begin{bmatrix} 0 & 10 & 5 & 10 \end{bmatrix}^\mathrm{T}$, $\mathbf{x}_{1|0} = \begin{bmatrix} 1 & 1 & 1 & 1 \end{bmatrix}^\mathrm{T}$, $\mathbf{P}_{1|0} = I_n$.

To comprehensively evaluate the performance of SR-GCI-IUKF in practical settings, experiments were conducted on a vehicle navigation system under various non-Gaussian noise environments. As shown in Figures 1-3, the proposed SR-GCI-IUKF achieves superior estimation accuracy for both position and velocity compared to all benchmark algorithms, with its parameter-optimized variant delivering the lowest RMSE. This performance gain directly demonstrates the efficacy of the GCI criterion in mitigating diverse complex noises and validates the critical role of the nonlinear error generalization model in correcting errors introduced by strong system nonlinearities. Furthermore, the consistent outperformance of the adaptive parameter strategy over fixed settings highlights its practical utility in automating the tuning process, which addresses a long-standing challenge in robust filter design. Consequently, these results substantiate that SR-GCI-IUKF successfully integrates and advances upon the core strengths of its predecessors—namely, the bandwidth insensitivity of CI and the numerical stability of square-root formulations—by introducing dynamic nonlinear correction and adaptive generalized kernel selection, thereby offering a more robust and practical solution for state estimation in challenging real-world applications.

To examine the parameter sensitivity of our proposed SR-GCI-IUKF algorithm, we conducted comparative experiments with deliberately varied parameter configurations under measurement noise e. As shown in Tables V and VI, the resulting ARMSE values demonstrate significant sensitivity to parameter selection. Notably, the parameter set (2.2, 20) and (2.3, 15) achieved the minimal ARMSE under noisy conditions, respectively, indicating its optimal configuration for this experimental scenario. This parameter sensitivity primarily stems from the inherent characteristics of the GCI mechanism, which enhances algorithm robustness at the cost of increased parametric complexity. Consequently, practitioners must perform systematic parameter tuning based on specific operational environments to optimize algorithmic performance. The experimental results underscore the importance of parameter adaptation to maintain superior filtering accuracy across diverse application scenarios.

### *C. Power system FASE*

This subsection evaluates the proposed algorithm's robustness in practical power system scenarios using the IEEE 14-bus test system.

The power system FASE is a tool for real-time estimation of power system state that makes full use of the priori information about the state to predict the system state and track changes in the node states. The state equation of the power system is as follows

$$\mathbf{x}_\tau = \mathbf{x}_{\tau-1} + \mathbf{w}_{\tau-1} \tag{95}$$

$$\mathbf{y}_\tau = \mathbf{h}(\mathbf{x}_\tau) + \mathbf{v}_\tau \tag{96}$$

where $\mathbf{x}_\tau = [a_{2,\tau}, a_{3,\tau}, \cdots, a_{L,\tau}, V_{1,\tau}, V_{2,\tau}, \cdots, V_{L,\tau}]^T \in \mathbb{R}^N$, containing voltage angle and voltage magnitude. Observation equation $\mathbf{h}(\bullet)$ is obtained from [9][29]

$$\begin{cases} J_y = \sum_{l=1}^{14} U_y U_l (G_{yy} \cos\theta_{yl} + B_{yl} \sin\theta_{yl}) \\ Q_y = \sum_{l=1}^{14} U_y U_l (G_{yl} \cos\theta_{yl} - B_{yl} \sin\theta_{yl}) \\ J_{y-l} = U_y{}^2 (G_{gy} + G_{yy}) - U_y U_l (G_{yl} \cos\theta_{yl} + B_{yl} \sin\theta_{yl}) \\ Q_{y-l} = -U_y{}^2 (B_{gy} + B_{yl}) - U_y U_l (G_{yl} \sin\theta_{yl} + B_{yl} \sin\theta_{yl}) \end{cases} \tag{97}$$

where $V_y$, $J_l$, $Q_{y-l}$ and $J_{y-l}$ are the voltage amplitude, the real power injection, the reactive power injection, the real power flow and the reactive power flow, respectively. $\theta_{yl}$ is the voltage angle between the bus bars *y* and *l*. $G_{yl}, G_{gy}, B_{yl}$ and $B_{gy}$ are the conductivity and the conductance.

The initial values of the states are obtained from [27]. Parameter configurations for all algorithms in this experiment are detailed in Table II.

*Case a. Gaussian Mixed Noise*

The robustness of our proposed algorithm in this case is tested by mixing two Gaussian noises with different variance sizes to simulate a non-Gaussian noise situation encountered in power systems. The tested noise is as follows

$$\mathbf{v}_\tau \sim 0.9 \times N(0, 0.01) + 0.1 \times N(0, 1000) \tag{98}$$

Fig. 4 compares the RMSE performance under Gaussian mixed noise within the highly nonlinear context of power system dynamic state estimation. The results demonstrate that the proposed SR-GCI-IUKF with parameter optimization consistently achieves the lowest RMSE. This superior performance stems from its dual capability: the GCI criterion robustly handles the non-Gaussian measurement outliers, while its integrated nonlinear error generalization model specifically addresses the state estimation errors induced by the strong nonlinearity of the power flow equations. This synergistic design ensures that SR-GCI-IUKF not only resists noise corruption but also maintains higher fidelity in tracking the true system state amidst nonlinear dynamics, where other filters exhibit accelerated performance degradation. Thus, the results validate the algorithm's practical advantage in realistic grid conditions characterized by both complex noise and inherent nonlinearity.

*Case b. Laplace Mixed Noise*

This subsection evaluates the stability of the proposed algorithm by testing it in a Laplace noise environment. The anomalous noise environments that may be encountered are modelled by mixing Laplacian and Gaussian noise, and the Laplacian mixing noise used is as follows

$$v_\tau \sim 0.9\times L(0,0.1)+0.1\times N(0,100) \tag{99}$$

where $L(0,0.1)$ and $N(0,100)$ denote Laplace noise with 0.1 mean and variance 1 and Gaussian noise with mean 0 and variance 100 respectively. Fig. 5 compares the RMSE performance under Laplace mixture noise, an environment characterized by heavier-tailed and more impulsive components than Gaussian noise. The results demonstrate that the proposed SR-GCI-IUKF with parameter optimization achieves superior accuracy, outperforming all benchmarked robust approaches. This pronounced advantage stems from the algorithm's tailored design: the GCI criterion inherently possesses greater flexibility in modeling heavy-tailed distributions compared to the Gaussian kernels used in MCC and MEE, allowing it to more effectively suppress the large errors induced by impulsive noise without over-rejecting valid measurements. Furthermore, the parameter adaptation strategy is crucial here, as it automatically adjusts the kernel's scale and shape to the specific intensity of the Laplace components, which is a non-trivial task for fixed-parameter filters like SR-CI-IUKF. Consequently, SR-GCI-IUKF not only proves robust to this challenging noise type but does so autonomously, validating its capability as a reliable and adaptive state estimator in systems prone to severe, non-Gaussian disturbances.

*Case c. Asymmetric Gaussian Noise*

In this subsection, we further evaluate the algorithm's robustness using a Gaussian hybrid noise model with distinct means and variances to emulate an asymmetric noise environment. The noise model used is as follows

$$v_\tau \sim 0.8\times N(0,1)+0.1\times N(1,100)+0.1\times N(-1,100) \tag{100}$$

Fig. 6 presents the results under asymmetric Gaussian noise. The proposed SR-GCI-IUKF maintains the lowest RMSE, demonstrating a distinct advantage in this scenario. The key differentiator is its ability to handle asymmetry—a challenge for filters with symmetric kernels such as MCC, MEE and CI. The GCI’s shape parameter and the adaptation strategy work synergistically to automatically characterize and compensate for the biased error distribution, rather than merely suppressing it. This confirms SR-GCI-IUKF’s unique suitability for real-world systems where sensor errors or disturbances are intrinsically non-symmetric.

*Case d. Computing time*

To evaluate the practical efficiency, the average computation time per iteration for each algorithm is compared in Table VII. The results indicate that the proposed SR-GCI-IUKF incurs a higher computational cost than the benchmarks. This overhead is a direct consequence of its advanced features: the square-root decomposition ensures numerical stability, the nonlinear augmented model dynamically corrects for strong nonlinearities, and the generalized Gaussian kernel operations underpin its robustness to complex noise. Crucially, the parameter adaptation strategy, while contributing to the time increase, is fundamentally what enables the algorithm to achieve robust performance without manual parameter fine-tuning across diverse environments. In practice, this overhead can be managed by narrowing the predefined parameter search space or triggering adaptation only when significant changes in innovation statistics are detected. Therefore, the increased computation time is not merely a drawback but a justifiable investment for obtaining superior estimation accuracy and robust autonomy in challenging, non-stationary scenarios where traditional filters fail or require extensive manual intervention.

## V. CONCLUSION

This paper proposes the SR-GCI-IUKF algorithm. The main contribution lies in proposing a robust UKF framework that is inherently insensitive to kernel bandwidth selection and capable of dynamically correcting errors induced by strong nonlinearities. This is achieved by integrating the GCI criterion, whose flexible kernel structure eliminates the critical bandwidth sensitivity of traditional Correntropy-based filters, and a nonlinear error generalization model that actively compensates for linearization inaccuracies during state updates. The framework is further stabilized by a square-root implementation, ensuring numerical robustness. Supported by an adaptive parameter strategy and theoretical stability analysis, the proposed algorithm provides a superior and practical solution for state estimation under complex non-Gaussian noise and nonlinear conditions, as validated by comprehensive simulations.

TABLE VII
COMPUTING TIME FOR DIFFERENT ALGORITHMS

| Algorithm | Computing time |
|---|---|
| UKF | 0.0071 |
| IUKF | 0.0101 |
| MCC-UKF | 0.0112 |
| MEE-UKF | 0.0150 |
| SR-CI-IUKF | 0.0197 |
| SR-GCI-IUKF(Trail) | 0.0200 |
| SR-GCI-IUKF (optimization) | 0.0807 |

## APPENDIX

First of all, we can easily get

$$\mathbf{F}_{\tau+1}\mathbf{K}_\tau\mathbf{V}_\tau\mathbf{K}_{\tau+1}^T\mathbf{F}_{\tau+1}^T \geq 0 \tag{101}$$

Therefore, we have

$$\mathbf{P}_{\tau+1|\tau} \geq \mathrm{M}_\tau\mathbf{P}_{\tau|\tau-1}\mathrm{M}_\tau^T+\mathbf{W}_\tau \tag{102}$$

Then according to the matrix inversion lemma, we can get

$$\mathbf{F}_{\tau+1}^{-1}\left(\mathbf{F}_{\tau+1}-\mathbf{F}_{\tau+1}\mathbf{K}_\tau\mathbf{H}_\tau\right)\mathbf{P}_{\tau|\tau-1}=\left(\mathbf{P}_{\tau|\tau-1}^{-1}+\mathbf{H}_\tau\mathbf{V}_\tau\mathbf{H}_\tau^T\right)^{-1} \tag{103}$$

Therefore, we can conclude that $\mathrm{M}_\tau$ is reversible. We have

$$\mathbf{P}_{\tau+1|\tau} \geq \mathrm{M}_\tau\left(\mathbf{P}_{\tau|\tau-1}+\mathrm{M}_\tau^{-1}\mathbf{W}_\tau\mathrm{M}_\tau^{-T}\right)\mathrm{M}_\tau^T \tag{104}$$

Then by ***Assumption 1***,

$$\mathrm{M}_\tau^{-1}=\left(\mathbf{F}_{\tau+1}-\mathbf{F}_{\tau+1}\mathbf{K}_\tau\mathbf{H}_\tau\right)^{-1} \geq \frac{1}{\bar{\mathrm{f}}+\bar{\mathrm{f}}\bar{\lambda}_{\mathbf{K}_\tau}\bar{\mathrm{h}}} \tag{105}$$

Therefore, (104) can be transformed into

$$\mathbf{P}_{\tau+1|\tau} \geq \mathrm{M}_\tau \left( \mathbf{P}_{\tau|\tau-1} + \frac{\underline{\mathrm{w}}}{\left(\bar{\mathrm{f}} + \bar{\mathrm{f}}\bar{\lambda}_{\mathbf{K}_\tau}\bar{\mathrm{h}}\right)^2} \right) \mathrm{M}_\tau^T \tag{106}$$

Taking the inverse matrix on both sides first, then multiplying left by $\mathrm{M}_\tau^T$ and right by $\mathrm{M}_\tau$ on the right, we get

$$\mathrm{M}_\tau^T \mathbf{P}_{\tau+1|\tau}^{-1} \mathrm{M}_\tau \leq \left( 1 + \frac{\underline{\mathrm{w}}}{\bar{\mathrm{p}}\left(\bar{\mathrm{f}} + \bar{\mathrm{f}}\bar{\lambda}_{\mathbf{K}_\tau}\bar{\mathrm{h}}\right)^2} \right) \mathbf{P}_{\tau|\tau-1}^{-1} \tag{107}$$

Therefore, we get $\gamma_\tau = 1 - \dfrac{1}{1 + \dfrac{\underline{\mathrm{w}}}{\bar{\mathrm{p}}\left(\bar{\mathrm{f}} + \bar{\mathrm{f}}\bar{\lambda}_{\mathbf{K}_\tau}\bar{\mathrm{h}}\right)^2}}$ in Lemma 2.